\documentclass[11pt,a4paper]{article}
\usepackage[english]{babel}
\usepackage{amsmath,amssymb,amsthm,epsfig,color,graphicx}
\usepackage[latin1]{inputenc}
\usepackage{xcolor}
\newtheorem{theorem}{Theorem}

\newtheorem{lemma}[theorem]{Lemma}
\newtheorem{definition}{Definition}
\newtheorem{corollary}{Corollary}

\definecolor{Red}{cmyk}{0,1,1,0}

\definecolor{Blue}{cmyk}{1,1,0,0}

\newcommand{\ba}{\begin{array}}
\newcommand{\ea}{\end{array}}
\newcommand{\be}{\begin{equation}}
\newcommand{\ee}{\end{equation}}
\newcommand{\ben}{\begin{enumerate}}
\newcommand{\een}{\end{enumerate}}

\let\g=\gamma

\let\s=\sigma

\let\L=\Lambda

\newcommand{\La}{\Lambda}

\begin{document}

\title{Phase Transitions in Ferromagnetic Ising Models with spatially dependent magnetic fields}
\author{Rodrigo Bissacot\\
\footnotesize{Applied Math Department}\\
\footnotesize{\texttt{rodrigo.bissacot@gmail.com}}\\
\footnotesize{Institut of Mathematics and Statistics - IME USP - University of S\~ao Paulo}
\\[0.3cm]
Marzio Cassandro\\
%\footnotesize{Mathematics Department}\\
\footnotesize{GSSI, Via. F. Crispi 7, 00167 L'Aquila, Italy}\\
\footnotesize{\texttt{marzio.cassandro@gmail.com}}
\\[0.3cm]
Leandro Cioletti\\
\footnotesize{Mathematics Department}\\
\footnotesize{Universidade de Bras\'ilia, 70910-900 Bras\'ilia - DF, Brasil}\\
\footnotesize{\texttt{leandro.mat@gmail.com}}\\
\\[0.1cm]
Errico Presutti\\
%\footnotesize{Dipartimento di Matematica}\\
\footnotesize{GSSI, Via. F. Crispi 7, 00167 L'Aquila, Italy}\\
\footnotesize{\texttt{errico.presutti@gmail.com}}\\
}
\maketitle

\begin{abstract}
In this paper we study the nearest neighbor Ising model with ferromagnetic interactions in the presence
of a space dependent  magnetic field which vanishes as $|x|^{-\alpha}$, $\alpha >0$, as $|x|\to \infty$.
We prove that in dimensions $d\geq 2$ for all $\beta$ large enough if $\alpha>1$ there is a phase transition
while if $\alpha<1$ there is a unique DLR state.

\end{abstract}

\section{Introduction}
\label{intro}
\hspace{0.46cm}
The Ising Model is one of the most studied subjects in Statistical Physics and will complete a century in a few years\footnote{Wilhelm Lenz introduced the model in 1920.}.
The literature about ferromagnetic Ising models on $\mathbb{Z}^d$, $d\ge 2$, is mainly focused on cases where the external magnetic field is constant. %; in general the field is equal to zero and we have phase transition.
We will study ferromagnetic nearest neighbor  hamiltonians of the form
\begin{eqnarray}
\label{hamilton}
H^{w}_{\L}(\s)= - J\sum_{|x-y|=1, x,y \in \L}\s(x)\s(y) - \sum_{x\in \L} h(x)\s(x)
- J \sum_{|x-y|=1, x \in \L, y \notin \L} \s(x)w(y)
\end{eqnarray}
where $\Lambda$ is any finite subset of $\mathbb{Z}^d$, $\s \in \{-1,1\}^{\Lambda}$
is a spin configuration in $\Lambda$, $w\in \{-1,1\}^{\Lambda^c}$ a boundary condition and $J >0$
the interaction strength.

When the magnetic field $h(\cdot)$  is constant, that is  $h(x)=h$  for
all $x \in \mathbb{Z}^d$ and $h=0$, then the classical Peierls' argument guarantees
the existence of a phase transition.  If instead $h \neq 0$  at all temperatures there is
a unique DLR measure, as it follows from the Lee-Yang Theory and GHS inequalities.
The absence of phase transitions comes from the differentiability
of the free energy with respect to the parameter $h$.

Alternating signs fields on the lattice $\mathbb{Z}^2$
are considered in \cite{NOZ},
constant fields on semi-infinite lattices are studied in \cite{Ba, FP}.
The magnetic field in all these models  has some spatial symmetry.
The challenging case of  i.i.d.\ random magnetic fields
on $\mathbb{Z}^d$  with zero mean has been studied in
\cite{AW,Bo,BK,COP1,COP2} and the case with positive mean in \cite{NF}.
Some deterministic and not spatially symmetric fields have
been considered in \cite{BC}.

In this paper we study the hamiltonian \eqref{hamilton} in $\mathbb{Z}^d$, $d\ge 2$,
with a non negative,
space dependent magnetic field $h(\cdot)$ of the form %, the prototype being
\begin{equation}
\label{0.1} 
h(x) = \begin{cases}\frac{h^* }{|x|^{\alpha}}& x\ne 0\\
h^* & x=0\end{cases},\quad \alpha >0, h^*>0
\end{equation}
where if $x=(x_1,\ldots,x_d)$ then  $|x|= \sum_{i=1}^{d}|x_i|$.
Calling $ Z^w_{\beta, h(\cdot),\L}$ the corresponding partition function
one can easily check that (along van Hove sequences)
\[
\lim_{\La\to \mathbb Z^d} \frac{\log Z^w_{\beta, h(\cdot),\L}}{\beta |\La|}
= p_{\beta}
\]
independently of the boundary conditions $w$.  The limit  $p_{\beta}$
is equal to the thermodynamic pressure without magnetic fields (i.e.\  $h^*=0$). This
indicates that
the presence of  $h(\cdot)$ does not change the thermodynamics thus suggesting
that a phase transition may occur for $\beta$ large, just as when the magnetic field is absent.
However surface effects are relevant
in the analysis of phase transitions and indeed we
shall prove in Theorem \ref{thm3.1}
that when $\alpha<1$ there is a unique
DLR measure, while
when $\alpha >1$ there is a phase transition for $\beta$ large enough,
see Theorem \ref{thm2.1}.

The existence of phase transitions at $\alpha>1$ is based on the validity of the
Peierls bounds for contours.  The proof of uniqueness when $\alpha<1$ at low  temperatures is more involved and it is based on an iterative scheme introduced in \cite{BMPZ}. For $\alpha = 1$ we have partial results but not a complete characterization.

\section{Existence of phase transitions}
\label{sec:2}

In this section we shall prove:

    \medskip

    \begin{theorem}
    \label{thm2.1}
Let $h(\cdot)$ be as in \eqref{0.1} with $\alpha>1$. Then for $\beta$ large enough there is
a phase transition, namely the  plus and minus Gibbs measures $\mu_{\beta, h(\cdot),\La}^{\pm}$
converge weakly as $\La\to \mathbb Z ^d$ to mutually distinct DLR measures.

     \end{theorem}

    \medskip

As we shall see the result extends to $\alpha=1$
under the additional assumption that $h^*$ is small enough
and to non negative  magnetic fields which are
``local perturbations'' of \eqref{0.1}
(by this we mean that
the $L^1$ norm of the difference is finite). We shall first prove the theorem under a stronger
assumption on the  magnetic field, see \eqref{2.5} below, which
allows to reproduce the Peierls' argument.  We need some geometric notation
that will be used extensively throughout the paper.

\begin{definition}
\label{defin2.1}
Two sites
$x$ and $y$ in $\mathbb Z^d$ are
connected iff they are nearest neighbors.
Given a finite set $K$  in $\mathbb Z^d$ we call $\bar K$ its complement,
$\delta_{\rm out}(K)$ the sites $y\in \bar K$ which are connected
to sites $x\in K$ and $\delta_{\rm in}(K)$ those in $K$ connected
to sites in $\bar K$.  $|\partial K|$ denotes the number of connected pairs $x,y$
with $x\in \delta_{\rm in}(K)$ and $y\in \delta_{\rm out}(K)$.

\end{definition}

\medskip

\begin{lemma}
\label{lemma2.3ME}

Let $h(\cdot)$ be any non negative magnetic field such that
\begin{equation}
    \label{2.5}
 J |\partial \Delta| > 2 \sum_{x\in \Delta}  h(x)
    \end{equation}
for all finite regions $\Delta \subset \mathbb{Z}^d$.
%($\partial \Delta$.  the  bonds from $\Delta$ to $\Delta^c$).
Then
for all $\beta$ large enough there is a phase transition.

\end{lemma}

\medskip

    \noindent
{\bf Proof.}
We shall use \eqref{2.5} to prove the validity
of the Peierls bounds, see \eqref{2.7} below. Then for all $\beta$ large enough
the weak limits of the Gibbs measures
with plus and minus boundary conditions are distinct DLR measures
$\mu_{\beta,  h(\cdot)}^{\pm}$. We thus have  a
phase transition hence the lemma.  We shall use later
that $\mu_{\beta,  h(\cdot)}^{\pm}$ have trivial $\s$-algebra
at infinity so that they have disjoint support, see for instance the Georgii book, \cite{G}.

Proof of the Peierls bounds.
Contours are geometric objects in the dual lattice $\mathbb Z^d_*$, namely call $C_x$,
$x\in \mathbb Z^d$, the closed unit cube in $\mathbb R^d$ with center $x$, then
 $\mathbb Z^d_*$ is the union over all n.n. pairs $x,y$ of the faces $C_x\cap C_y$.
 Given a spin configuration $\s$ its contours $\g$ are the maximal connected (in the sense
 of non void intersection) components of the union of all faces  $C_x\cap C_y$
 with $\s(x)\ne\s(y)$. 

Let $\g$ be a contour and
$I(\g)$ the interior of $\g$, i.e.\ the points which are connected to $\infty$ only via paths which
cross $\g$.  Suppose $\g$ is a minus contour  i.e.\ $\s(y)=-1$ on
$\delta_{\rm out}(I(\g)$).
%and call  $\partial \g$ the sites in $I(\g)$
%which are connected to
%$I(\g)^c$.
Denote by $Z_{I(\g);h(\cdot)}^-
(\s_{I(\g)}(x)=1, x\in \delta_{\rm in}(I(\g)) $ the
partition function in $I(\g)$ with magnetic field $h(\cdot)$,
minus boundary conditions and with the constraint that
$\s_{I(\g)}(x)=1$ for all $ x\in \delta_{\rm in}(I(\g)$.
Then
\begin{eqnarray*}
&& Z_{I(\g);h(\cdot)}^-
(\s_{I(\g)}(x)=1, x\in \delta_{\rm in}(I(\g)))
\\&& \hskip2cm \le
 e^{\beta \sum_{x\in I(\g)} h_x}\;
Z_{I(\g); h\equiv 0}^-(\s_{I(\g)}(x)=1, x\in \delta_{\rm in}(I(\g)))
\\&& \hskip2cm \le
e^{-2\beta J |\partial I(\g)|} e^{\beta \sum_{x\in I(\g)} h_x} Z_{I(\g); h\equiv 0}^-(\s_{I(\g)}(x)=-1, x\in \delta_{\rm in}(I(\g)))
\\&& \hskip2cm \le
e^{-2\beta J |\partial I(\g)|} e^{2\beta \sum_{x\in I(\g)} h_x}
Z_{I(\g);h(\cdot)}^-(\s_{I(\g)}(x)=-1, x\in \delta_{\rm in}(I(\g))).
    \end{eqnarray*}
Thus by \eqref{2.5} the weight of the contour $\g$ is bounded by
    \begin{equation}
    \label{2.7}
    \frac{Z_{I(\g);h(\cdot)}^-
(\s_{I(\g)}(x)=1, x\in \delta_{\rm in}(I(\g)))}
{Z_{I(\g);h(\cdot)}^-(\s_{I(\g)}(x)=-1, x\in \delta_{\rm in}(I(\g)))} \le
    e^{-\beta J |\partial I(\g)|}.
    \end{equation}
Same bound holds for the plus contours.  \qed

\medskip

The proof of Theorem \ref{thm2.1} will be obtained by
reducing to magnetic fields for which \eqref{2.5} is satisfied,
a task that will be achieved via a few lemmas where we
shall extensively use
the Isoperimetric Inequality  (see \cite{LP} for a proof):
for any finite $\Delta\subset \mathbb{Z}^d$ $(d\geq 2)$
\[
|\Delta|^{\frac{d-1}{d}}\leq \frac{|\partial \Delta|}{2d}.
\]

\medskip

\begin{lemma}
\label{lemma2.1ME}
Let $h(\cdot)$ be as in \eqref{0.1} with $\alpha>1$.  Then there is $C\equiv C(h^*,\alpha,d,J)>0$ so that
\eqref{2.5} holds for all finite regions $\Delta$ such that $|\Delta|>C$.

\end{lemma}
\medskip

\noindent{\bf Proof.}
Since $h(x)$ is a non
increasing function of $|x|$, calling
${B}(0,R):= \{ x: |x| \le R\}$  we have
\[
\sum_{x\in \Delta}  h(x) \le \sum_{x\in {B}(0,R)}h(x),\quad \text{ for $R$ such that }\;  |{B}(0,R)| \ge|\Delta|
\]

%($+1$ because $h(0)=0$).
We claim that
the condition $ |{B}(0,R)| \ge|\Delta|$ is satisfied
if
	\begin{equation}\label{raio}
		R= \text{ smallest integer } \ge c |\partial \Delta|^{\frac{1}{d-1}}
	\end{equation}
with $c$ large enough.  In fact, recalling that
$|\partial {B}(0,n)| = 2d\cdot n^{d-1}$, we have
$|{B}(0,R)| \ge a R^d$, $a>0$ small enough, hence using the isoperimetric inequality
\[
|{B}(0,R)| \ge a R^d \ge ac^d |\partial \Delta|^{\frac{d}{d-1}}
\ge ac^d (2d)^{\frac{d}{d-1}}|\Delta|  \ge|\Delta|
\]
for $c$ large enough.

Thus the lemma will be proved once we show that
\[
\lim_{R\to \infty} \frac 1{R^{d-1}} \sum_{|x|\le R} h(x)=0.
\]
Recalling that
$|\partial {B}(0,n)| = 2d\cdot n^{d-1}$ this is implied by
\[
\lim_{R\to \infty}\sum_{n=1}^{R} \frac {n^{d-1}}{R^{d-1}}\frac 1{n^{\alpha}} =0
\]
whose validity follows from the Lebesgue dominated
convergence theorem. The lemma is thus proved.
\qed

\vskip.5cm

Observe that  when $\alpha =1$ and $h^*$ is small enough then
\eqref{2.5} holds again
for all finite regions $\Delta$  large enough.  The proof is analogous except at the end as we only have
\[
\limsup_{R\to \infty} \frac 1{R^{d-1}} \sum_{|x|\le R} \frac 1{|x|^{\alpha}}\le c
\]

\vskip.5cm

\begin{lemma}
\label{lemma2.2ME}

Let $h(\cdot)$ be as in \eqref{0.1} with $\alpha>1$, then
there is $R$ so that \eqref{2.5}
holds for all finite $\Delta$ when the magnetic field is $\hat h$:
\[
\hat h (x)= \begin{cases} 0 & \text{if $|x|\le R$}\\
h(x)&\text{if $|x|> R$}\end{cases}
\]

\end{lemma}

\medskip

\noindent
{\bf Proof.}  Suppose  $|\Delta| > C$,  $C$
the constant in Lemma \ref{lemma2.1ME}, then
\[
2\sum_{x\in \Delta}  \hat h(x) \le
2\sum_{x\in \Delta}  h(x) \le J |\partial \Delta|
\]
Suppose next $|\Delta| \leq C$, then by the Isoperimetric Inequality,
\begin{eqnarray*}
\sum_{x\in \Delta}  \hat h(x) &=& \sum_{x\in \Delta; |x| > R}  \hat h(x)
\\ &\leq &   \frac{h^*|\Delta| }{R^{\alpha}} \leq  \frac{h^*|\partial \Delta|^{\frac{d}{d-1}} }{R^{\alpha}(2d)^{\frac{d}{d-1}}}
 \leq  \frac{h^*C^{\frac{1}{d-1}}|\partial \Delta| }{R^{\alpha}(2d)^{\frac{d}{d-1}}}
\end{eqnarray*}
which is $\le J|\partial \Delta|$ for  $R$ sufficiently large.
\qed

\vskip.5cm

\medskip

    \noindent
{\bf Proof of Theorem  \ref{thm2.1}.} Let $h(\cdot)$ be as in \eqref{0.1} with $\alpha>1$. By Lemma \ref{lemma2.3ME} and \ref{lemma2.2ME} for $\beta$ large enough
there is a phase transition for the system with magnetic field $\hat h(\cdot)$,
let $\mu_{\beta, \hat h(\cdot)}^{\pm}$ the corresponding DLR measures obtained as
limit of the Gibbs measures with plus respectively minus boundary conditions.
Call $ \phi(x):=  h(x) -\hat h(x)= \mathbf 1_{|x|<R} h(x)$
and define the probability measures
    \begin{equation}
    \label{2.6}
 d\nu_{\beta, h(\cdot)}^{\pm}(\s) := C_{\pm}e^{\beta \sum \phi(x) \s(x)}
d \mu_{\beta, \hat h(\cdot)}^{\pm}(\s)
    \end{equation}
($C_{\pm}$ the normalization constants).  
We shall first check that they are DLR measures with magnetic field $h(\cdot)$. To have lighter notation we drop super and subscripts writing just $\nu$, $\mu$ and $C$.
We need to show that for any finite cube $\La$ large enough
(we need below that $\La \supset B(0,R)$) the $\nu$ conditional probability
given $\s_{\bar \La}$ is the Gibbs measure with magnetic field $h(\cdot)$.  By the DLR property for $\mu$ we have
\[
d\nu(\s)= Ce^{\beta \sum \phi(x) \s(x)}
\frac{ e^{-\beta \hat H(\s_\La|\s_{\bar \La})}}{\hat Z_\La(\s_{\bar \La})}
d\mu_{\bar\La}(\s_{\bar \La})
\]
where $d\mu_{\bar\La}(\s_{\bar \La})$ is the marginal of $\mu$ on the
spin configurations in $\bar \La$.  We then have
\[
d\nu(\s)= C
\frac{ e^{-\beta  H(\s_\La|\s_{\bar \La})}}{  Z_\La(\s_{\bar \La})}
\frac{ Z_\La(\s_{\bar \La})}{\hat Z_\La(\s_{\bar \La})}
d\mu_{\bar\La}(\s_{\bar \La})
\]
By integrating over $\s_\La$ we get
\[
d\nu_{\bar\La}(\s_{\bar\La})= C
\frac{ Z_\La(\s_{\bar \La})}{\hat Z_\La(\s_{\bar \La})}
d\mu_{\bar\La}(\s_{\bar \La})
\]
hence
\[
d\nu(\s)=
\frac{ e^{-\beta  H(\s_\La|\s_{\bar \La})}}{  Z_\La(\s_{\bar \La})}
d\nu_{\bar\La}(\s_{\bar\La})
\]
which proves the DLR property.
Thus $ d\nu_{\beta, h(\cdot)}^{\pm}(\s)$
are DLR measures with magnetic field $h(\cdot)$ and  are absolutely
continuous w.r.t.\ $\mu_{\beta, \hat h(\cdot)}^{\pm}$. Hence they also have
disjoint supports and are therefore distinct. Theorem  \ref{thm2.1} is proved.  \qed

\section{Restricted ensembles and contour partition functions}
\label{sec:3}
We fix hereafter $h(x)$ as in \eqref{0.1}
and  we shall prove that
    \medskip

    \begin{theorem}
    \label{thm3.1}

Let $h(\cdot)$ as in \eqref{0.1}, then for any $\beta$ large enough there is a unique  DLR measure.

     \end{theorem}
\vskip.5cm

In this section we shall prove some crucial estimates
which will be used in the next section to  prove Theorem \ref{thm3.1}
but which have an interest in their own right.
Observe that when $h(\cdot)$ is given by
\eqref{0.1} the condition \eqref{2.5} may fail
for some $\Delta$ for instance a large ball centered at the origin.

With this in mind we classify the contours $\g$ by saying that $\g$ is ``slim'' if
   \begin{equation}
    \label{b3.2}
 J |\partial I(\g)| > 2 \sum_{x\in I(\g)}  h(x)
    \end{equation}
 see the proof of Lemma \ref{lemma2.3ME} for notation.
%
%where $I(\g)$ is the interior of $\g$, namely the union of all sites $x$ such that
%if a path connects $x$ to infinity then necessarily it  crosses $\g$.
We call ``fat'' the contours which do not satisfy \eqref{b3.2}.  Following Pirogov-Sinai
we then introduce plus-minus
restricted ensembles where spin configurations are restricted in such a way that
there are only slim contours.  We thus define for any bounded region $\La$     the plus-minus restricted partition functions
   \begin{equation}
    \label{b3.3}
Z^{\pm,{\rm slim}}_\La : =  \sum_{\s_\La:\text{all contours are slim}} e^{-\beta H(\s_\La | \pm \mathbf 1_{\La^c})}.
    \end{equation}
Obviously the pressures in the plus and  minus ensembles are equal but the Pirogov-Sinai
theory requires for the existence of a phase transition
finer conditions on the finite volume corrections to the pressure namely that the latter differs from the limit
pressure by a surface term.  In our case the correction is larger than a surface term
because $\alpha<1$ as
shown by the following:

\medskip

    \begin{theorem}
    \label{thmb3.2}

For any $\beta$ large enough there are positive constants $c_1$ and $c_2$ so that
   \begin{equation}
    \label{b3.4}
Z^{-,{\rm slim}}_\La \le c_1 e^{-\beta c_2 \sum_{x\in \La} h(x)} Z^{+,{\rm slim}}_\La.
    \end{equation}
     \end{theorem}

     \medskip
     \noindent
{\bf Proof.}     By repeating the proof of Theorem \ref{thm2.1} and denoting by $E^{-,{\rm slim}}_\La$
the expectation w.r.t.\ the Gibbs measure in the minus restricted ensemble,
we have for any $x\in \La$:
   \begin{equation}
    \label{b3.5}   
E^{-,{\rm slim}}_\La (\s(x))\le -1 +   2 \sum_{\g: I(\g)\ni 0} e^{-\beta J|\partial I(\gamma)|} = -m^*,\quad m^*>0
    \end{equation}
for $\beta$ large enough.  Then
   \begin{equation}
    \label{b3.6}
\mu_{\beta, h(\cdot),\La}^{-,{\rm slim}} \Big[ \frac{\sum_{x\in \La} h(x)\s_\La(x)}{\sum_{x\in \La} h(x)} \le -\frac{m^*}2\Big]
\ge  \frac{m^*}{2-m^*} %e^{-\beta m^* \sum_{x\in \La} h(x)}
    \end{equation}
To prove \eqref{b3.6} let $X$ be a random variable with values in $[-1,1]$ and $P$ its law.
Suppose that $E(X) \le - m^*$ and call  $p:= P[X\ge - m^*/2]$, then
   \begin{equation*}
%    \label{b3.6}
-m^* \ge -1(1-p) -\frac{m^*}2 p,\quad (1-\frac{m^*}2)p \le (1-m^*),\quad (1-p)  \ge
\frac{m^*}{2-m^*}
    \end{equation*}
hence \eqref{b3.6}.

Calling $Z^{-,{\rm slim}}_\La(A)$ the partition function with the constraint $A$,
we can rewrite \eqref{b3.6} as:
   \begin{eqnarray*}
Z^{-,{\rm slim}}_\La  &\le&  \frac{2-m^*}{m^*} Z^{-,{\rm slim}}_\La\Big( \frac{\sum_{x\in \La} h(x)\s_\La(x)}{\sum_{x\in \La} h(x)} \le -\frac{m^*}2\Big)\\
 &\le&  \frac{2-m^*}{m^*} e^{-\beta \frac{m^*}2\sum_{x\in \La} h(x)}
 Z^{-,{\rm slim}}_{\La, h\equiv 0}
 \\
 &=&  \frac{2-m^*}{m^*} e^{-\beta \frac{m^*}2\sum_{x\in \La} h(x)}
 Z^{+,{\rm slim}}_{\La, h\equiv 0}.
    \end{eqnarray*}
By repeating the previous argument we get
   \begin{equation*}
 Z^{+,{\rm slim}}_{\La, h\equiv 0} \le  \frac{2-m^*}{m^*} e^{-\beta \frac{m^*}2\sum_{x\in \La} h(x)}
  Z^{+,{\rm slim}}_{\La}
    \end{equation*}
where $  Z^{+,{\rm slim}}_{\La}$ is the partition function with
the contribution of the magnetic field
$h(\cdot)$. This  concludes the proof of the theorem.
\qed

\vskip.5cm
In the next section we shall use a corollary of Theorem \ref{thmb3.2} that we
state after  introducing some notation. The geometry is as  follows:

$\La$ is a cube with center the origin,
$\Delta$ a subset of $\La$ and $K$ a subset of $\Delta$ which is union
of disjoint connected set $K_i$ where for each $i$ the complement $\bar K_i$ of $K_i$ has
a unique maximally connected component (i.e.\ there are no ``holes'' in $K_i$). We also
suppose that each $K_i$ is fat, i.e.\ {\color{red}}
   \begin{equation*}
 J |\partial K_i|  \le   2 \sum_{x\in K_i}  h(x)
    \end{equation*}
    and that
$\delta_{\rm out} K \subset \Delta$, see \  Definition \ref{defin2.1}.
%
%
% where:
%given a set $A$ we denote by
%$\delta_{\rm out} A$ the set of all $x\in \bar A$ which are connected to $A$ and by
%$\delta_{\rm in} A$ the set of all $x\in  A$ which are connected to $\bar A$.

With $\La$,
$\Delta$ and $K$ as above we denote by \  $\mathcal X_{\La,\Delta,K,M}$,
$M \subset \delta_{\rm out} \Delta$,
the set of
all configuration $\s_\La$ which have the following properties.

\begin{itemize}

\item  $\s_\La =-1$ on $\delta_{\rm in} \Delta$, $\s_\La =-1$ on $M \subset \delta_{\rm out} \Delta$
and  $\s_\La =+1$ on $\delta_{\rm out} \Delta\setminus M$.

\item  $\s_\La =-1$ on $\delta_{\rm  } K$
and  $\s_\La =+1$ on $\delta_{\rm in} K$.

\item  $\s_\La$ has only slim contours in $ \Delta\setminus K$

\end{itemize}

\noindent
We denote by $Z_\La^\omega(\mathcal X_{\La,\Delta,K,M})$ the partition function in $\La$
with constraint $\mathcal X_{\La,\Delta,K,M}$ and boundary conditions $\omega$.  Then:

\medskip

    \begin{corollary}
    \label{corob3.1}

Under the same assumptions of Theorem \ref{thmb3.2}
   \begin{equation}
    \label{b3.6.1}
Z_\La^\omega(\mathcal X_{\La,\Delta,K,M}) \le c_1 e^{-\beta c_2 \sum_{x\in \Delta\setminus K} h(x)}
e^{-2\beta J |\partial K |}e^{ -2\beta J |\partial\Delta| + 4\beta J |M|} Z_\La^\omega
    \end{equation}
     \end{corollary}

     \medskip
In the applications of the next section the connected
components of $\Delta$  should intersect some given set and this will enable to control
the sum over $\Delta$ via the bound $e^{ -2\beta J |\partial \Delta|}$.
 
  The sum over $K$
is instead controlled as follows.
We  introduce the fat-contours partition function on the whole $\mathbb Z^d$ as
   \begin{equation}
    \label{b3.7}
Z^{{\rm fat}} : =  \sum_{n=0}^\infty \sum^*_{\g_1,..,\g_n} e^{-\beta J \sum |\partial I(\g_i)|}
    \end{equation}
where the sum $*$ refers to a sum over only fat contours   such that $I(\g_i)\cap
I(\g_j)=\emptyset$
for all $i\ne j$.

\medskip

    \begin{theorem}
    \label{thmb3.3}

For any $\beta$ large enough there is a positive constant $c_3$   so that
   \begin{equation}
    \label{b3.8}
Z^{{\rm fat}} \le c_3
    \end{equation}
     \end{theorem}

     \medskip
     \noindent
{\bf Proof.}  We order the points of $\mathbb Z^d$ in a way which respects the distance from the origin
and given a contour $\g$ we denote by $X(\g)$ the minimal point in $\g$ with the given order.  By
the definition of fat contours and supposing $X(\g)\ne 0$,
  \begin{equation*}
%    \label{b3.8}
J|\partial I(\g)| \le 2 \sum_{x\in I(\g)} h(x) \le \frac{2h^*}{|X(\g)|^\alpha} |I(\g)|
\le \frac{2h^*C_p}{|X(\g)|^\alpha}|\partial I(\g)|^{\frac d{d-1}}
    \end{equation*}
where $C_p$ is the isoperimetric constant. Hence
  \begin{equation}
    \label{b3.9}
|\partial I(\g)| \ge  (\frac{J}{2C_p h^*})^{d-1} |X(\g)|^{\alpha(d-1)},\quad X(\g)\ne 0
    \end{equation}
We write
   \begin{eqnarray*}
Z^{{\rm fat}} &=& \sum_n\sum_{x_1,..,x_n} \sum^*_{\g_1,..,\g_n} \prod_{i=1}^n
\mathbf 1_{X(\g_i)=x_i}e^{-\beta J  |\partial I(\g_i)|}\\&\le &
\prod_{x\in \mathbb Z^d}\Big (1+ \sum_{\g \; {\rm fat}: X(\g)=x} e^{-\beta J|\partial I(\g)|}\Big )
\\&= & (1+ \sum_{\g \; {\rm fat}: X(\g)=0} e^{-\beta J|\partial I(\g)|}\Big )
\prod_{x \ne 0}\Big (1+ \sum_{\g \; {\rm fat}: X(\g)=x} e^{-\beta J|\partial I(\g)|}\Big )
    \end{eqnarray*}
which using \eqref{b3.9} proves \eqref{b3.8}.
\qed

\vskip1cm

Before moving to the next section with the proof of Theorem \ref{thm3.1} we
point out that by the Dobrushin's Uniqueness Theorem  there is a unique DLR state also at high temperatures and since the system is ferromagnetic, uniqueness may be expected
to hold at
all temperatures.
However the proof of such a statement when the external field is zero does not
seem to extend easily to our case, see \cite{CMR} and \cite{Hag}.

\section{Uniqueness at low temperatures}
\label{sec:4}

In this section we prove Theorem \ref{thm3.1}.  For any  positive integer $n$ we denote
by $\La_n$ the cube with center the origin
and side $2n+1$.
We fix a positive integer $L$, eventually $L\to \infty$, and  arbitrarily the spins outside $\La_{L}$, denoting by $\mu_L$
the Gibbs measure on $\{-1,1\}^{\La_{L}}$ with the given boundary conditions
and external magnetic field as in \eqref{0.1}.

\medskip

\noindent
{\bf Definitions.}
\begin{itemize}

\item Given $\s_{\La_L}$,   $\Delta \subset \La_L$, $B: B\cap \Delta=\emptyset$
we say
that
 $x\in \Delta$  is $-$ connected in $\Delta$ to $B$ if there is $X\subset \Delta$ such that: $x\in X$,
$X$ is connected to $B$ and $\s_\La\equiv -1$ on $X$.

\item

Let $\mathfrak{C}_L$ be the random set of sites  $x\in \La_{L}$
which are $-$ connected in $\La_{L}$ to $\La_{L+1}\setminus \La_L$ and let $\mathfrak{M}_k =
\mathfrak{C}_L \cap \La_{k+1}\setminus \La_k$, $k<L$;  $\mathfrak{M}_L= \La_{L+1}\setminus \La_L$.

\item    Given $k\le L$ and $M\subset \La_{k+1}\setminus \La_k$ we
define   $\mathfrak{C}_{k,M}(\s_{\La_L})$ as the set of all $x\in \La_k$ which are
$-$ connected in $\La_k$ to $M$.  In particular $\mathfrak{C}_{L,M}=\mathfrak{C}_{L}$
if $M=\La_{L+1}\setminus \La_L$.

%
%
%If $C$ is a set  we denote by $\delta_{\rm out}(C)$ the sites in the complement $\bar C$
%of $C$
%which are connected to $C$. % and which can be connected to $\infty$ without crossing $C$.

\end{itemize}

\medskip

\noindent
Suppose $\mathfrak{C}_L=C$ then the spins in

$\delta_{\rm out}(C\cup \bar\La_L)$ are all equal to $+1$.
Moreover if we change the configuration $\s_\La$ leaving unchanged
the spins in $C':=(C\cup \bar\La_L) \cup \delta_{\rm out}(C\cup \bar\La_L)$ we still have $\mathfrak{C}_L=C$.

Thus the spins in $\La_L\setminus C'$ are distributed
with Gibbs measure with plus boundary conditions.    We shall prove that
there exists $b^*<1$ so that
    \begin{equation}
    \label{3.2}
\lim_{L\to \infty} \mu_{L}\Big[ \mathfrak{C}_L \cap \La_{L(1-b^*)} =\emptyset\Big]=1
    \end{equation}
which then proves that $\mu_{L}$ converges weakly to the plus DLR measure, which is
the weak limit of Gibbs measures with plus boundary conditions.
Thus any DLR measure is equal to the plus DLR measure and  Theorem \ref{thm3.1}
is proved.  We are therefore reduced
to the proof of \eqref{3.2} which
uses an iterative argument introduced in \cite{BMPZ}.

%
%\medskip
%
%\noindent
%{\bf Definition.}

\medskip

%\noindent
It readily follows from the definitions that for $k<L$:
    \begin{equation}
    \label{b4.3}
\mathfrak{C}_L \cap \La_{k} = \mathfrak{C}_{k,\mathfrak{M}_k}, \;\;\;\;    \mathfrak{M}_k=
\mathfrak{C}_L \cap ( \La_{k+1}\setminus \La_k).
    \end{equation}
The next  property will be used to establish a connection with  Corollary \ref{corob3.1}, it
is therefore crucial in the proof of Theorem  \ref{thm3.1}. We claim that:
    \begin{equation}
    \label{b4.3nn}
\s_{\La_L}(x) = 1\, \text{ for all $x$ in $\delta_{\rm out}(\mathfrak{C}_{k,\mathfrak{M}_k})\setminus \mathfrak{M}_k$}
    \end{equation}
Proof: By definition $\s_{\La_L}(x) = 1$ for all $x$ as in \eqref{b4.3nn} which
are in $\La_k$.  It remains to consider all $x$ as in \eqref{b4.3nn} which are in
$\La_{k+1}\setminus \La_k$.
%
%in\cap \delta_{\rm out}(\mathfrak{C}_{k,\mathfrak{M}_k})\setminus \mathfrak{M}_k$.
%
We argue by contradiction supposing $\s_{\La_L}(x)=-1$.  In such a case there is a
path with all minuses which starts at $x$ and ends in $\mathfrak{M}_k$.  Since $\mathfrak{M}_k \subset \mathfrak{C}_L$ and $\mathfrak{C}_L$ is connected,
then $x\in \mathfrak{C}_L$ which implies (since $x\in\La_{k+1}\setminus \La_k$) that $x\in \mathfrak{M}_k$, hence the contradiction.  \eqref{b4.3nn} is proved.

Before proceeding we need some extra notation:

{\bf Notation.} We decompose $\mathfrak{C}_{k,M}$ into maximally connected components, each one of them
is a connected set whose complement has an unbounded maximally connected component
and maybe several maximally connected finite components.  The latter are distinguished
into fat and slim and we call $\bar {\mathfrak{C}}_{k,M}^{\rm fat}$ and $\bar {\mathfrak{C}}_{k,M}^{\rm slim}$  the union of all
the fat, respectively slim ones.

\medskip

It then follows directly from \eqref{b4.3nn} that
    \begin{equation}
    \label{b4.3z}
 \{   \mathfrak{M}_k = M \} \cap \{  \bar {\mathfrak{C}}_{k,M}^{\rm fat} = K  \} \cap
  \{ \mathfrak{C}_{k,M} \cup \bar {\mathfrak{C}}_{k,M}^{\rm slim} = \Delta  \}
 \subset \mathcal X_{\La,\Delta,K,M}
    \end{equation}
    $\mathcal X_{\La,\Delta,K,M}$ the set considered in Corollary \ref{corob3.1}.

We are now ready for the proof of Theorem \ref{thm3.1}.
The basic point is that if $|\mathfrak{M}_{k_0}|$ is small for some $k_0$
then (with large probability) there is $k>k_0$ with $|\mathfrak{M}_{k}|$
even smaller. Iterating the argument we will then find a $k$ where $|\mathfrak{M}_{k}|=0$.
The heuristic idea behind the proof of such properties is the following.

Suppose that  $|\mathfrak{M}_{k_0}|= L^a$, $a>0$, $k_0$ a fraction of $L$. Let $0<a'<a$,
fix a constant $b<1$  suitably small and distinguish two cases:
    \[
 |\mathfrak{M}_k| \le L^{a'}\quad \text{for some $k\in [ k_0-bL,k_0)$}
\]
and the complement where
\begin{equation}
    \label{b4.3nnn}
 |\mathfrak{M}_k| > L^{a'}\quad \text{for all $k\in [k_0-bL,k_0)$}
\end{equation}
 We argue that the event \eqref{b4.3nnn} has vanishing  probability
as $L\to \infty$.  To this end we  use \eqref{b4.3z} (with $k=k_0$) and  Corollary \ref{corob3.1} observing
that (with the above notation) $|\Delta| \ge |\mathfrak{C}_{k_0,M}| \ge bL L^{a'}$,
by \eqref{b4.3nnn}.
In \eqref{b3.6.1} we then have a dangerous term $e^{4\beta J  L^a}$
(which comes from $M =\mathfrak{M}_{k_0}$, $|M|\le L^a$), while the contribution
of the magnetic field is bounded by $e^{ -\beta c_2 (h^*L^{-\alpha})bL L^{a'}}$.  If
\[
L^{1+a' -\alpha} > L^{a}
\]
the magnetic field wins against the dangerous term.  We need a lengthy counting
argument to sum over all possible values of $\Delta$, $K$ and $M$ which will be given
in the end of the section and which will prove  that
with probability going to 1 as $L\to \infty$ we can  reduce to the case
$|\mathfrak{M}_k| \le L^{a'}$ for some $k\in [ k_0-bL,k_0)$.

We can satisfy
the previous inequality with
$a' = a - \frac{1-\alpha}2$ and then iterate the argument to prove that
after finitely many steps we get  $\mathfrak{M}_{k}=\emptyset$
and thus conclude the proof.

%
%
%In the latter set $\mathfrak{C}_{L}$  has cardinality larger than
%$bL L^{a'}$.
%{\color{red}
%As we shall see using \eqref{b4.3} and \eqref{b4.3nn} we are in the setup of Corollary \ref{corob3.1} with $|\Delta\setminus K| \ge |\mathfrak{C}_{L}| \ge bL L^{a'}$
%and $M=\mathfrak{M}_{k_0}$, $|M| \le L^a$.  It will then follow using Theorem \ref{thmb3.3} that the probability of the
%event in \eqref{b4.3nnn} is bounded proportionally to
%$e^{4\beta J  L^a} e^{ -\beta c_2 (h^*L^{-\alpha})bL L^{a'}}$. If
%\[
%L^{1+a' -\alpha} > L^{a}
%\]
%then the bound vanishes as $L\to \infty$
%} so that
%with probability going to 1 as $L\to \infty$ we can  reduce to the case
%$|\mathfrak{M}_k| \le L^{a'}$ for some $k\in [ k_0-bL,k_0)$.  We can satisfy
%the previous inequality with
%$a' = a - \frac{1-\alpha}2$ and then iterate the argument to prove that
%after finitely many steps we get  $\mathfrak{M}_{k}=\emptyset$
%and thus conclude the proof.

With this in mind we introduce the sequence $a_n$, $n\ge 0$,
by setting
    \begin{equation}
    \label{3.3}
a_0=d-1,\;\; a_{n+1}=a_n - \frac{1-\alpha}2
    \end{equation}
and call $n^*$ the largest integer such that $a_{n^*}\ge 0$.
Let $s_0=L$ and define recursively $s_n$
for $1\le n\le n^*$ by setting
    \begin{equation}
    \label{3.4}
\text{$s_n$ the largest $k$ not larger than $s_{n-1}$ such that $|\mathfrak{M}_k| \le L^{a_n}$}
    \end{equation}
and, if there is no  $k$ as in \eqref{3.4},  we then set $s_n=0$ and  stop the sequence.
Observe that if $|\mathfrak{M}_{s_{n-1}}|=0$ then $s_n=s_{n-1}$.   If not stopped earlier we define $s_{n^*+1}$ as
    \begin{equation}
    \label{3.5}
\text{$s_{n^*+1}$ is the largest $k$ not larger than $s_{n^*}$ such that $|\mathfrak{M}_k| =0$}
    \end{equation}
setting $s_{n^*+1}=0$ if $k$ does not exist.

Let $b>0$ be such that
    \begin{equation}
    \label{3.6}
b n^* < \frac 1{100}
    \end{equation}
Then $\mathfrak{C}_L \cap \La_{L(1-b^*)} =\emptyset$ in the set
    \begin{equation}
    \label{3.7}
\mathcal G := \bigcap_{1\le n\le n^*+1} \{ s_{n-1}-s_n \le bL\}
    \end{equation}
provided $b^*>1/2$ so that \eqref{3.2}  will follow once we prove  that
    \begin{equation}
    \label{b4.8}
\lim_{L\to \infty} \mu_{L}\Big[ \mathcal G\Big]=1.
    \end{equation}

We shall prove that for any $1\le p\le n^*+1$
    \begin{equation}
    \label{4.8}
\lim_{L\to \infty} \mu_L\Big[ s_{p+1} < s_{p}-bL \; ;\; s_{p} \ge L-pbL\Big]=0
    \end{equation}
which yields \eqref{b4.8}.

We write $\mu_L\big[ s_{p+1} < s_{p}-bL \; ;\; s_{p} \ge L-pbL\big]$
as the ratio
of two partition functions, the one in the denominator is the full partition
function $Z_{\La_L}^\omega$, $\omega$ the boundary conditions outside $\La_L$,
while the one in the numerator will be simply called $Z$ and it will be the object
of our analysis.  We decompose the configurations according to the value $k$ of
$s_p$ and $M$ of $\mathfrak{M}_k$. 
If $|M|=0$ we do not have to prove anything so that in the sequel we tacitly suppose
 $|M|>0$. We have
    \begin{eqnarray}
    \label{4.9}
Z &\le & \sum_{L\ge k \ge L-pbL} \;\;\;\sum_{M\subset \La_{k+1}\setminus \La_k: |M|\le L^{a_p}} \sum_{C_{k,M}\subset \La_k: |C_{k,M}|\ge  bL^{1+a_{p+1}}} \nonumber\\ && \hskip3cm
Z_{\La_L}\Big(\mathfrak{M}_k = M;\mathfrak{C}_{k,M}=C_{k,M}\Big)
    \end{eqnarray}
The sets $K =\bar {\mathfrak{C}}_{k,M}^{\rm fat}$ and $\bar {C}_{k,M}^{\rm slim} = \bar {\mathfrak{C}}_{k,M}^{\rm slim}$ are uniquely determined by $C_{k,M}$ and
we can rewrite \eqref{4.9} as
    \begin{eqnarray}
    \label{4.9bis}
Z &\le & \sum_{L\ge k \ge L-pbL} \;\;\;\sum_{M\subset \La_{k+1}\setminus \La_k: |M|\le L^{a_p}} \sum_{K,\Delta : |\Delta|\ge  bL^{1+a_{p+1}}} \nonumber\\ && \hskip1cm
Z_{\La_L}\Big(\mathfrak{M}_k = M; \bar{\mathfrak{C}}_{k,M}^{\rm fat} = K;
\mathfrak{C}_{k,M}\cup\bar {C}_{k,M}^{\rm slim}=\Delta\Big)
    \end{eqnarray}
observing that  $\Delta\subset \La_k$ is the union of a finite number
of disjoint connected sets (without ``holes'', see Section \ref{sec:3}), say $\Delta_1$,..,$\Delta_n$, each one  connected to $M$.  $K$ is the union of
fat connected sets without holes each one contained in  $\Delta$.  When we add a $*$ to
the sum over $K$ and $\Delta$ we mean that the sum is over sets with
such a restriction.  We then get from \eqref{4.9bis} after using \eqref{b4.3z} and
\eqref{b3.6.1}
    \begin{eqnarray}
    \label{4.9t}
Z &\le & \sum_{L\ge k \ge L-pbL} \;\;\;\sum_{M\subset \La_{k+1}\setminus \La_k: |M|\le L^{a_p}} \sum^*_{K,\Delta : |\Delta\setminus K|\ge  bL^{1+a_{p+1}}} \nonumber\\ && \hskip1cm
c_1 e^{-\beta c_2 h^* L^{-\alpha} bL^{1+a_{p+1}}}
e^{-2\beta J |\partial K |}e^{ -2\beta J |\partial \Delta| + 4\beta J |M|} Z_{\La_L}^\omega
    \end{eqnarray}
where $Z_{\La_L}^\omega$ is the full partition function.
We next specify the maximal connected components of $\Delta$, called $\Delta_1,..,\Delta_n$,
and use Theorem \ref{thmb3.3} and \eqref{b3.8} to perform the sum over $K$ then getting
    \begin{eqnarray}
    \label{4.9tt}
\frac{Z}{Z_{\La_L}^\omega} &\le & \sum_{L\ge k \ge L-pbL} \;\;\;\sum_{M\subset \La_{k+1}\setminus \La_k: |M|\le L^{a_p}} \sum_{n\ge 1}\sum^*_{\Delta_1,..,\Delta_n } \nonumber\\ && \hskip1cm
c_1 e^{-\beta c_2 h^* L^{-\alpha} bL^{1+a_{p+1}}}
 c_3^ne^{ -2\beta J |\partial \Delta| + 4\beta J  L^{a_p}}
    \end{eqnarray}
where the $*$ recalls that $\Delta_1,..,\Delta_n$ are mutually disjoint connected sets without holes
each one connected to $M$, this implies that the sum is over $n \le |M|\le L^{a_p}$.
 Each  $\Delta_i$ is then in one to one correspondence with $\delta_{\rm out}(\Delta_i)$,
which is a $*$connected set which intersects $M$.

 Thus
we can bound
the $*$ sum  by summing over $n\le |M|$ disjoint $*$connected sets which intersect $M$.
Hence
    \begin{equation}
    \label{4.13}
  \sum_{n\ge 1}\sum^*_{\Delta_1,..,\Delta_n }
e^{ -2\beta J |\partial \Delta|} \le \sum_{n=1}^{|M|} \frac{M!}{n!(M-n)!} e^{-\beta c_4 n}
\le \Big(1+e^{-\beta c_4}\Big)^{|M|}
     \end{equation}
where $c_4$ is such that
    \begin{equation}
    \label{4.14}
e^{-\beta c_4 } \ge\;\;\; \sum_{D \ni 0, D * {\rm connected}}\;\;\;  e^{ -2\beta J |D|}
     \end{equation}
\eqref{4.14} holds for $\beta$ large enough, see for instance Lemma 3.1.2.4 in 
 \cite{presutti}.

Then recalling \eqref{4.9tt}
    \begin{eqnarray}
    \label{4.14t}
\frac{Z}{Z_{\La_L}^\omega} &\le & \sum_{L\ge k \ge L-pbL} \;\;\;\sum_{M\subset \La_{k+1}\setminus \La_k: |M|\le L^{a_p}}  \nonumber\\ && \hskip1cm
c_1 e^{-\beta c_2 h^* L^{-\alpha} bL^{1+a_{p+1}}}
 c_3^{ L^{a_p}}e^{  4\beta J  L^{a_p}} \Big(1+e^{-\beta c_4}\Big)^{L^{a_p}}
    \end{eqnarray}
We can now perform the sum over $M$ which using the Stirling formula is bounded by
$ e^{ c_5 L^{a_p} \log L}$, $c_5$ a suitable constant    and thus get
        \begin{eqnarray}
    \label{4.14tt}
\frac{Z}{Z_{\La_L}^\omega} &\le &  L  e^{ c_5 L^{a_p} \log L}
c_1 e^{-\beta c_2 h^* L^{-\alpha} bL^{1+a_{p+1}}}
 c_3^{ L^{a_p}}e^{  4\beta J  L^{a_p}} \Big(1+e^{-\beta c_4}\Big)^{L^{a_p}}
    \end{eqnarray}
 %
%
%
%
%\newpage
%
%    \begin{eqnarray*}
%\frac{Z}{Z_{\La_L}}
%&\le&
%c_1c_3e^{-\beta (c_2 bL^{1+a_{p+1}-\alpha} -4JL^{a_p})} \Big(1+e^{-\beta c_4}\Big)^{L^{a_p}}
%L e^{ c_5 L^{a_p} \log L}
%    \end{eqnarray*}
which recalling the definition of $a_n$
proves that
    \begin{equation}
    \label{4.15}
\mu_L\Big[ s_{p+1} < s_{p}-bL \; ;\; s_{p} \ge L-pbL\Big] \le c_6 e^{-\beta \frac{c_2}2 bL^{1+a_{p+1}-\alpha}}
    \end{equation}
    thus proving \eqref{4.8} and hence \eqref{b4.8}.

\vskip1cm

\section{Concluding remarks}
\label{sec:5}

We have proved that when the magnetic field is given by \eqref{0.1}
for all $\beta$ large enough there is a phase transition when $\alpha >1$ while, if $\alpha<1$, there is a unique DLR state. It seems plausible that uniqueness extends to all $\beta$ but we do not have a proof. Using the random cluster representation uniqueness is related to the absence of percolation (see \cite{CMR}), perhaps this can be useful to deal with this question.
 When $\alpha =1 $ and $h^*$ small enough
the proof of Section \ref{sec:2} applies and we thus have a phase transition.
However, our proof of uniqueness does not extend to the case
$\alpha=1$ no matter how large is $h^*$ and a different approach should be used
maybe related to an extension of Minlos-Sinai or the Wulff shape problem.\\

\noindent{\bf Acknowledgments}\\

\noindent The authors thank Aernout van Enter for fruitful discussions. Rodrigo Bissacot is supported by the Grant 2011/22423-5 from FAPESP and Grant 308583/2012-4 from CNPq. Leandro Cioletti is supported by FEMAT.  The authors thank  Maria Eul\'{a}lia Vares and the organizers of the XVII Brazilian School of Probability  during which the $\alpha >1$ case was proved. Rodrigo Bissacot acknowledges very kind hospitality and support from GSSI  in L'Aquila, and from the Mathematics Department of the University of Bras\'{i}lia.

\end{document}